\documentclass[conference]{IEEEtran}
\IEEEoverridecommandlockouts
\usepackage{cite}
\usepackage{amsmath,amssymb,amsfonts,booktabs}
\usepackage{algorithm, algorithmic}
\usepackage{graphicx}
\usepackage{bm}
\usepackage{textcomp}
\usepackage{xcolor}
\usepackage{subcaption}
\usepackage{multirow}
\usepackage{siunitx}
\usepackage[stretch=40,shrink=40]{microtype}

\usepackage[normalem]{ulem}

\newcommand{\real}{\mathbb{R}}          
\newcommand{\trans}{^{\text{T}}}		

\def\BibTeX{{\rm B\kern-.05em{\sc i\kern-.025em b}\kern-.08em
    T\kern-.1667em\lower.7ex\hbox{E}\kern-.125emX}}
\begin{document}

\title{Efficient Convolutional Forward Modeling and Sparse Coding in Multichannel Imaging\\
\thanks{This work was partially supported by the Fraunhofer Internal Programs under Grant No. Attract 025-601128, and the  Thuringian Ministry of Economic Affairs, Science and Digital Society (TMWWDG), as well as by Amazon Fund.}
}

\author{\IEEEauthorblockN{Han Wang\textsuperscript{*,\ddag}, Yhonatan Kvich\textsuperscript{\dag}, Eduardo P\'{e}rez\textsuperscript{*,\ddag}, Florian R\"{o}mer\textsuperscript{*,\ddag}, Yonina C. Eldar\textsuperscript{\dag}}
\IEEEauthorblockA{\textsuperscript{*}\textit{Applied AI Signal Processing and Data Analysis, Fraunhofer Institute for Nondestructive Testing}, Saarbrücken, Germany \\
\textsuperscript{\dag}\textit{Faculty of Math and Computer Science, Weizmann Institute of Science}, Rehovot, Israel \\
\textsuperscript{\ddag}\textit{Dept. Electronic Measurements and Signal Processing, Technische Universit\"{a}t Ilmenau}, Ilmenau, Germany\\
\{han.wang, eduardo.jose.perez.mejia, florian.roemer\}@izfp.fraunhofer.de, \{yonatan.kvich, yonina.eldar\}@weizmann.ac.il}}

\maketitle

\begin{abstract}
This study considers the Block-Toeplitz structural properties inherent in traditional multichannel forward model matrices, using Full Matrix Capture (FMC) in ultrasonic testing as a case study. We propose an analytical convolutional forward model that transforms reflectivity maps into FMC data. Our findings demonstrate that the convolutional model excels over its matrix-based counterpart in terms of computational efficiency and storage requirements. This accelerated forward modeling approach holds significant potential for various inverse problems, notably enhancing Sparse Signal Recovery (SSR) within the context LASSO regression, which facilitates efficient Convolutional Sparse Coding (CSC) algorithms. Additionally, we explore the integration of Convolutional Neural Networks (CNNs) for the forward model, employing deep unfolding to implement the Learned Block Convolutional ISTA (BC-LISTA). 
\end{abstract}


\section{Introduction}\label{sec_introduction}
When tackling inverse problems to obtain descriptive parameters based on observations, the solution approach is directly related to the forward model describing the observed data \cite{palomar2010convex}. In applications such as ultrasound nondestructive testing and medical imaging, as well as radar localization, the prior knowledge of structural sparsity has led to the usage of algorithms such as Orthogonal Matching Pursuit (OMP) \cite{GOMP, bossmann2012sparse}, Iterative Shrinkage-Thresholding Algorithm (ISTA) \cite{ISTA}, and its accelerated variant, Fast ISTA (FISTA) \cite{FISTA}, which require iterative optimization of forward model parameters. Recent strides in model-based deep learning \cite{shlezinger2022model} have given rise to algorithms like Learned ISTA (LISTA) \cite{LISTA, monga2021algorithm}, reducing the number of iterations in inference, and to task-based compressed sensing applications, ranging from optimal sensor placement to sophisticated subsampling techniques \cite{wangeusipco2023, mulleti2023learning, wangius2023, dps, wangicassp2024}. Despite their innovations, traditional matrix-based forward models, alongside Multilayer Perceptron (MLP)-based LISTA implementations, grapple with the computational demands imposed by burgeoning data volumes.

In response, deep learning-based solutions \cite{wei2018deep, chen2020review} and matrix-free forward models \cite{matrix_free_1, matrix_free_2} have emerged as innovative alternatives. Among these, convolutional forward models, which utilize specific matrix structures to facilitate Convolutional Sparse Coding (CSC) algorithms, represent a significant advancement. Notably, previous studies have proposed convolutional dictionaries as effective substitutes for matrix-based models with a Toeplitz structure \cite{sreter2018learned}. Furthermore, the introduction of a LISTA-Toeplitz model has demonstrated the potential of replacing Gram matrices appearing in the gradient of linear least squares problems, which have a Toeplitz structure in some applications, with convolutional filters, though the approach still harbors a significant count of trainable parameters due to its reliance on linear layers for representing the remaining components of the gradient \cite{fu2021structured}. Nonetheless, such assumptions do not always align with the realities of multichannel time-delay-based measurements, where instead a more general Block-Toeplitz structure is present \cite{kailath1994generalized, semper2018defect}.

Recognizing this gap, we demonstrate that the inherent structure of Uniform Linear Array (ULA) measurements allows the forward model to be precisely rewritten using ensembles of strided convolutions. Our main contributions are as follows: first, we develop an accurate convolutional forward model by leveraging the block Toeplitz structure of the physical model matrix. Second, we introduce the Block Convolutional FISTA (BC-FISTA) and Block Convolutional LISTA (BC-LISTA) algorithms for signal recovery. BC-FISTA is shown to be mathematically equivalent to the naive FISTA implementation but requires significantly less memory. BC-LISTA enables the learning of convolutional kernels and hyperparameters, providing enhanced flexibility and improved performance. Third, our comparative analysis reveals that the convolutional model significantly outperforms traditional matrix-based models regarding memory efficiency. Furthermore, both proposed CSC algorithms are capable of generating high-quality images.


\section{Matrix-Based Forward Model}\label{sec_image_formation}
\begin{figure}[ht!]
	\centering
    \vspace{-0.3cm}
	\includegraphics[width=0.95\columnwidth, trim = {0cm 0cm 0cm 0.3cm}, clip]{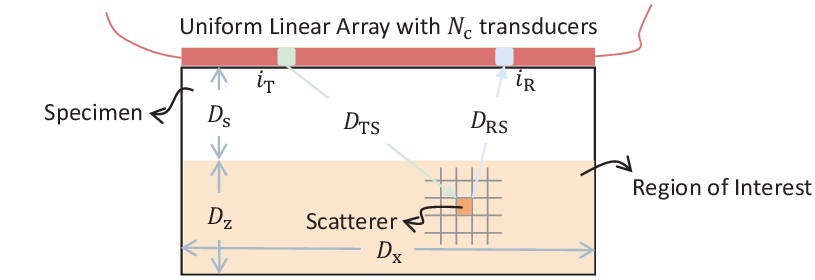}
	\caption{Two-dimensional illustration of FMC measurement.}
    \vspace{-0.2cm}
	\label{fig_FMC}
\end{figure}
Consider a ULA comprising $N_{\rm{c}}$ transducers with a pitch size of $d_{\rm{c}}$, strategically positioned on the specimen targeted for detection. Each transducer in this array has dual functionality, capable of both transmitting and receiving signals. The total numbers of transmitters and receivers are respectively denoted by $N_{\rm{T}}$ and $N_{\rm{R}}$. There is a distance $D_{\rm{s}}$ between the array and the Region of Interest (ROI) to avoid artifacts. The ROI is of shape $(D_{\rm{z}}\times D_{\rm{x}})$ and discretized into $(N_{\rm{z}}\times N_{\rm{x}})$ pixels, each measuring $(d_{\rm{z}}\times d_{\rm{x}})$ in size. Each pixel is a potential scatterer that can reflect the pulse to receivers. 

In the experiment, the Full Matrix Capture (FMC) refers to sequentially excite element by element, and all receivers capture the response signals for each transmission. We assume that the received signal $p(t)$ at a given receiver constitutes a superposition of time-delayed, amplitude-scaled replicas of the emitted pulse, each subject to phase inversion and attributable to the interaction with discrete scatterers. Mathematically, the received signal is given by the real-valued Gabor function:
\begin{equation}\label{pulse_shape}
    p(t)=\sum_{s=1}^{N_{\rm{s}}} a_{s}e^{-\alpha(t-\tau_{s})^2}\cos(2\pi f_c (t-\tau_{s})) + n(t),
\end{equation}
where $s$ and $N_{\rm{s}}$ are the index and total number of scatterers within the ROI, respectively. The variable $\tau_s$ signifies the Time of Flight (ToF) and is a function of the spatial coordinates of the transmitter, the receiver, and the individual scatterer. The parameter $a_s$ represents the reflectivity coefficient associated with each scatterer, encapsulating the relative strength of the echo. The term $\alpha$ is indicative of the bandwidth factor, which, along with the center frequency $f_c$, characterizes the pulse shape. The additional term $n(t)$ accounts for Additive White Gaussian Noise (AWGN) introduced during signal reception. Given the sampling frequency $f_{\rm{s}}$ and duration, $N_{\rm{t}}$ samples are obtained per A-scan. Finally, we assemble a three-dimensional measurement data denoted as $\mathbf{Y}\in \mathbb{R}^{N_{\rm{t}} \times N_{\rm{R}} \times N_{\rm{T}}}$

We characterize the ROI by a two-dimensional reflectivity map $\mathbf{X}\in \mathbb{R}^{N_{\rm{z}}\times N_{\rm{x}}}$. The conventional approach for constructing the forward model commences with the vectorization of the measurement data and the reflectivity map, adhering to the Fortran-style (column-major) format. Subsequently, each column of the model matrix $\mathbf{A}$ is designated to represent the anticipated FMC dataset that would result from a scatterer located at the respective coordinate. In this context, the forward model $\mathbf{A}\in \mathbb{R}^{{(N_{\rm{t}}N_{\rm{R}}N_{\rm{T}}}) \times (N_{\rm{z}}N_{\rm{x}})}$ is formulated. This model maps the vectorized reflectivity map $\mathbf{x} \in \mathbb{R}^{N_{\rm{z}}N_{\rm{x}}}$ to the vectorized FMC measurement data $\mathbf{y} \in \mathbb{R}^{N_{\rm{t}}N_{\rm{R}}N_{\rm{T}}}$. Then this measurement procedure is effectively expressed as a linear transformation:
\begin{equation}\label{eq_FMC}
    \mathbf{y=Ax+n},
\end{equation}

The matrix-based forward model (\ref{eq_FMC}) encounters severe limitations regarding storage capacity and computational efficiency, especially with larger ULAs and wider ROIs. The convolutional forward model presents a viable solution to these challenges, which we introduce next.

\section{Convolutional Forward Model}\label{sec_conv_model}
\subsection{Block-Toeplitz Structure Forward Model}
\begin{figure}[t!]
	\centering
	\includegraphics[width=0.9\columnwidth, trim = {0cm 0cm 0cm 0cm}, clip]{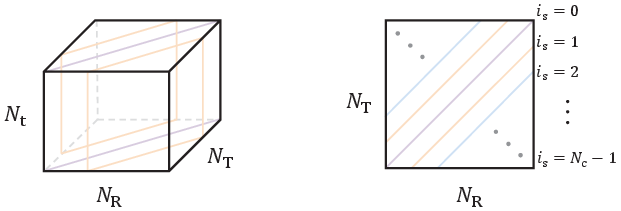}
	\caption{Slices of 3D FMC data array.}
    \vspace{-0.4cm}
	\label{fig_FMC_data_slices}
\end{figure}

As depicted in Fig. \ref{fig_FMC_data_slices}, the multi-dimensional FMC measurement data $\mathbf{Y}$ can be conceptualized as comprising $(2N_{\rm{c}}-1)$ diagonal slices. Owing to the reciprocity principle between transmitters and receivers, two slices positioned symmetrically about the center slice exhibit identical characteristics. Consequently, the data $\mathbf{Y}$ can be effectively represented by only $N_{\rm{c}}$ unique slices. We denote a specific 2D data slice by $\mathbf{Y}_{i_{\rm{s}}}$, where $i_{\rm{s}}$ ranges from $0$ to $N_{\rm{c}}-1$, and each slice contains $(N_{\rm{s}}-i_{\rm{s}})$ A-scans. Each A-scan is associated with a unique transmitter-receiver pair, with the relationship between their indices formulated as $i_{\rm{R}}=i_{\rm{T}}+i_{\rm{s}}$. To elucidate further, within the $i_{\rm{s}}$ slice, the index $i_{\rm{T}}$ spans the set $[0, ..., N_{\rm{c}}-1-i_{\rm{s}}]$, while $i_{\rm{R}}$ correspondingly traverses the range $[i_{\rm{s}}, ..., N_{\rm{c}}-1]$.

\begin{figure*}[ht!]
	\centering
	\includegraphics[width=0.75\textwidth, trim = {0cm 0cm 0cm 0cm}, clip]{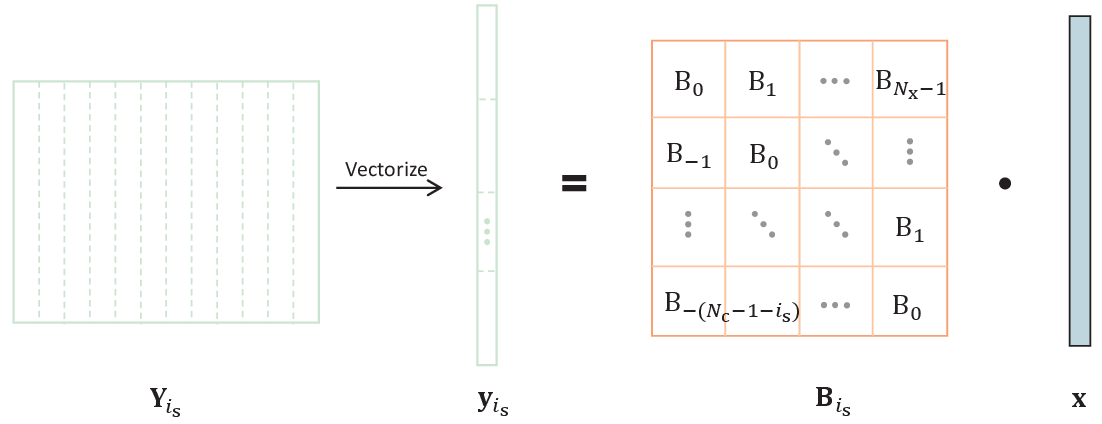}
	\caption{Slice-wise matrix-based forward model.}
	\label{fig_slice_model}
\end{figure*}

In the context of the slice-wise matrix-based forward model illustrated in Fig. \ref{fig_slice_model}, the matrix $\mathbf{B}_{i_{\rm{s}}}\in \real^{(N_{\rm{c}}-i_{\rm{s}})N_{\rm{t}}\times (N_{\rm{z}}N_{\rm{x}})}$ can derive the vectorized slice $\mathbf{y}_{i_{\rm{s}}}$ from the reflectivity map $\mathbf{x}$. The matrix $\mathbf{B}_{i_{\rm{s}}}$ is intricately structured into $(N_{\rm{c}}-i_{\rm{s}})\times N_{\rm{x}}$ blocks. Each block encapsulates the signal determined by the specific coordinates of transmitters, receivers, and scatterers:
\begin{itemize}
    \item Transmitters: $[i_{\rm{T}}\cdot d_{\rm{c}}, \, 0]$  
    \item Receivers: $[(i_{\rm{T}}+i_{\rm{s}})\cdot d_{\rm{c}}, \, 0]$, where $i_{\rm{T}}\in [0, ..., N_{\rm{c}}-1-i_{\rm{s}}]$
    \item Scatterers: $[i_{\rm{x}}\cdot d_{\rm{x}}, \, i_{\rm{z}}\cdot d_{\rm{z}}+D_{\rm{s}}]$, where $i_{\rm{x}}\in [0, ..., N_{\rm{x}}-1]$ and $i_{\rm{z}}\in [0, ..., N_{\rm{z}}-1]$.
\end{itemize}
Using these coordinates and assuming that the pixel width is equal to the pitch size, such that $d_x=d_c$, the ToF is computed via the equation: 
\begin{equation}\label{eq_tof2}
\begin{split}
    \tau = \frac{1}{c_0}\bigg( &\sqrt{[(i_{\rm{T}}-i_{\rm{x}})d_{\rm{x}}]^2+(i_{\rm{z}}d_{\rm{z}}+D_{\rm{s}})^2} \\
    &+\sqrt{[(i_{\rm{T}}+i_{\rm{s}}-i_{\rm{x}})d_{\rm{x}}]^2+(i_{\rm{z}}d_{\rm{z}}+D_{\rm{s}})^2}\bigg).
\end{split}
\end{equation}

It is important to observe that the ToF parameter, $\tau$, remains invariant for constant values of the difference $(i_{\rm{T}}-i_{\rm{x}})$. The constancy leads to a situation where the blocks along any given diagonal of the matrix $\mathbf{B}_{i_{\rm{s}}}$ are identical. Consequently, Fig. \ref{fig_slice_model} demonstrates the Block-Toeplitz structure inherent in the slice-wise matrix-based forward model. The matrix $\mathbf{B}_{i_{\rm{s}}}$ is amenable to a more compact representation, utilizing only $(N_{\rm{c}} - i_{\rm{s}} + N_{\rm{x}} - 1)$ unique blocks to encapsulate the entire structure. This tailored configuration implies that the forward model can be efficiently distilled into a series of one-dimensional convolutions, significantly simplifying computational efforts.

\subsection{Convolutional Model for Block-Toeplitz Matrix Structure}

\begin{figure*}[ht!]
	\centering
	\includegraphics[width=0.9\textwidth, trim = {0cm 0cm 0cm 0cm}, clip]{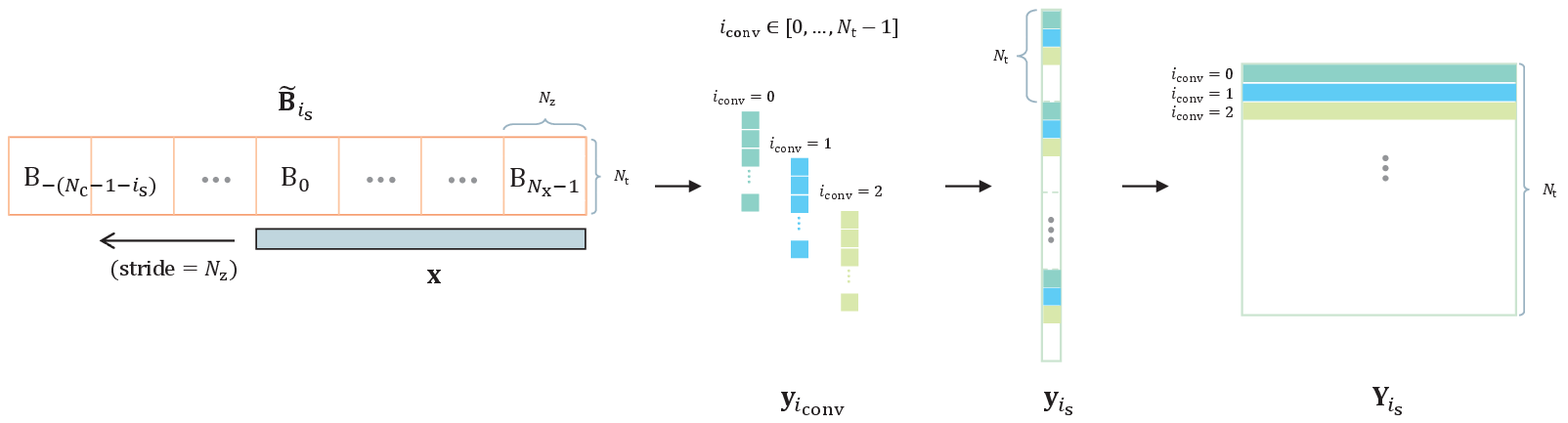}
	\caption{{Matrix-vector product from Fig. \ref{fig_slice_model} rewritten exactly as 1D strided convolutions.} \iffalse{1D convolutional model for Block-Toeplitz matrix.}\fi}
    \vspace{-0.3cm}
	\label{fig_1d_conv_model}
\end{figure*}

To reformulate the matrix-vector multiplication in the Block-Toeplitz structured matrix as convolution operations, the unique blocks are first reoriented and restructured into an elongated matrix $\tilde{\mathbf{B}}_{i_{\rm{s}}}\in \real^{N_{\rm{t}}\times (N_{\rm{c}} - i_{\rm{s}} + N_{\rm{x}} - 1)N_{\rm{z}}}$. Subsequently, executing a row-wise inversion of the matrix $\tilde{\mathbf{B}}_{i_{\rm{s}}}$ is essential to aligning the convolution kernels with the standard convolution operation paradigm. This adjustment allows for the vector $\mathbf{x}$ to be subject to convolution with each flipped row, utilizing a stride of $N_{\rm{z}}$. Each individual one-dimensional convolution operation yields a vector $\mathbf{y}_{i_{\text{conv}}}\in \real^{N_{\rm{c}}-i_{\rm{s}}}$. Upon completion of $N_{\rm{t}}$ such convolutions, the corresponding row vectors are concatenated in sequence to construct the data slice $\mathbf{Y}_{i_{\rm{s}}}$, a process depicted in Fig. \ref{fig_1d_conv_model}. 

Transitioning to a CNN framework for the forward model offers compelling advantages such as expedited inference, and efficient backpropagation. This CNN-based methodology can be smoothly integrated with other network architectures or algorithms, thereby enriching its utility and operational effectiveness. To accommodate the CNN structure, it is necessary to append zeros to either end of the input vector $\mathbf{x}$, with the padding determined by $(N_{\rm{c}} - i_{\rm{s}} - 1)N_{\rm{z}}$ and maintaining a stride of $N_{\rm{z}}$. This zero-padding preprocessing step does not impinge upon computational efficiency as the convolution count and kernel dimensions remain unaltered. By setting the number of output channels to $N_{\rm{t}}$, the output is congruent with $\mathbf{Y}_{i_{\rm{s}}}$, thereby obviating the need for iterative convolutions and manual assembly of the row vectors.

\begin{figure}[t!]
	\centering
	\includegraphics[width=0.95\columnwidth, trim = {0cm 0cm 0cm 0cm}, clip]{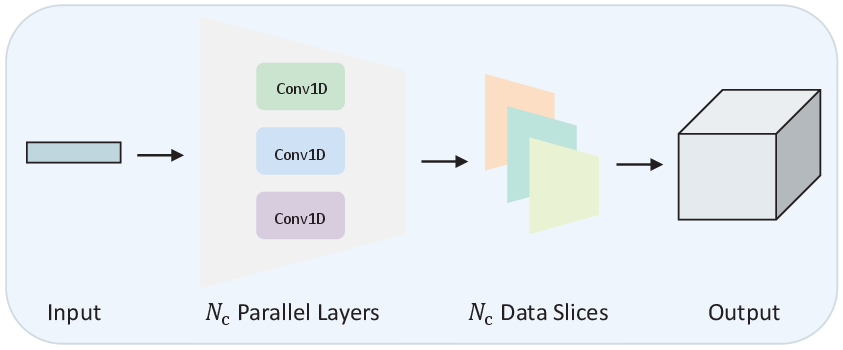}
	\caption{CNN-based forward model.}
    \vspace{-0.5cm}
	\label{fig_cnn_forward_model}
\end{figure}

In our proposed methodology, we instantiate $N_{\rm{c}}$ parallel Conv1D layers within our model architecture, with each layer responsible for the computation of a distinct data slice. These computed slices are then meticulously integrated into their corresponding positions to construct the FMC data array, which is visually delineated in Fig. \ref{fig_cnn_forward_model}. To summarize, the analytic convolutional forward model comprises $N_{\rm{c}}$ different matrices of varying dimensions, denoted $[\tilde{\mathbf{B}}_0, \tilde{\mathbf{B}}_1, ..., \tilde{\mathbf{B}}_{N_{\rm{c}}-1}]$. Each matrix serves to initialize the weights of a Conv1D layer configured with $N_{\rm{t}}$ output channels.

\begin{figure}[ht!]
	\centering
	\includegraphics[width=0.9\columnwidth, trim = {0cm 0cm 0cm 0cm}, clip]{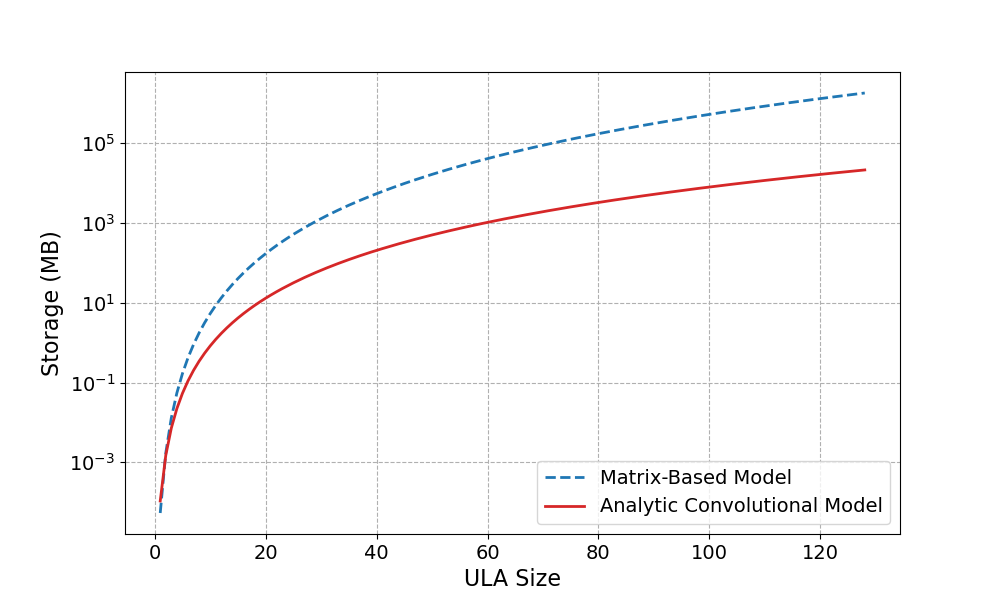}
	\caption{Model storage requirement comparison.}
 \vspace{-0.3cm}
	\label{fig_storage_compare}
\end{figure}

The analytical convolutional forward model exhibits several advantages over the matrix-based model, with the number of parameters serving as a pivotal aspect. To illustrate a quantitative evaluation of their respective performances, we consider a ULA with channel number, $N_{\rm{c}}$, ranging from $1$ to $128$. For simplicity, we set $N_{\rm{x}}=N_{\rm{z}}=N_{\rm{c}}$, $d_{\rm{x}}=d_{\rm{z}}=d_{\rm{c}}$, and $D_{\rm{s}}=D_{\rm{z}}$. The total number of samples, $N_{\rm{t}}$, is dictated by the longest ToF within the ROI. By assuming a single-precision float data type for the parameters, we calculate and compare the storage demands of both models in Megabytes, as shown in Fig. \ref{fig_storage_compare}. The convolutional model is obviously storage-friendly especially when the array size is larger. Consider a commonly-used $128$-channel ULA as an example, under this scenario, the matrix-based forward model requires approximately $1789\, \text{GB}$. In stark contrast, the convolutional forward model necessitates merely $21\, \text{GB}$. 

\section{Efficient Convolutional Sparse Coding}\label{sec_eff_csc}
The efficiency of our convolutional forward modeling facilitates effective gradient computation for backward passes, making it particularly suited for SSR. This capability underpins the development of efficient CSC algorithms for solving the LASSO function:
\begin{equation}\label{lasso}
    \min_{\mathbf{x}}\frac{1}{2}\Vert f(\mathbf{x})-\mathbf{y} \Vert_2^2 + \lambda \Vert \mathbf{x} \Vert_1,
\end{equation}
where $f(\mathbf{x})$ is the forward model, $\mathbf{y}$ is the observed data, $\mathbf{x}$ is the signal to be recovered, and $\lambda$ is a regularization parameter that controls the sparsity of the solution.

\subsection{Block Convolutional FISTA}
\begin{algorithm}[ht]
  \renewcommand{\algorithmicrequire}{\textbf{Input:}}
  \renewcommand{\algorithmicensure}{\textbf{Output:}}
  \caption{Block Convolutional FISTA}
  \label{alg_convfista}
  \begin{algorithmic}[1]
    \REQUIRE Initial guess $\mathbf{x}_0$,  auxiliary sequence $\mathbf{z}_0=\mathbf{x}_0$, Nesterov momentum $t_0=1$, number of iterations $N_{\text{iter}}$ \\
    \FOR{$i$ to $N_{\text{iter}}$}
        \STATE Compute gradient of $\frac{1}{2}\Vert f(\mathbf{z}_k)-\mathbf{y} \Vert_2^2$, denoted $\nabla f(\mathbf{z}_k)$
        \STATE Perform a gradient descent step: \\
                $\mathbf{x}_k=\mathbf{z}_k-\frac{1}{L}\nabla f(\mathbf{z}_k)$   
        \STATE Apply the proximal operator corresponding to the $L1$ regularization term to $\mathbf{x}_k$: \\
               $\mathbf{x}_{k+1} = S_{\lambda / L}(\mathbf{x}_k)$
        \STATE $S_{\theta}(v)$ is the soft-thresholding operator defined as: \\
               $S_{\theta}(v) = \text{sign}(v) \max(|v| - \theta, 0)$
        \STATE Update the momentum term $t$: \\
               $t_{k+1}=\frac{1+\sqrt{1+4t_k^2}}{2}$
        \STATE Update the auxiliary variable $\mathbf{z}$: \\
               $\mathbf{z}_{k+1}=\mathbf{x}_{k+1}+\frac{t_k-1}{t_{k+1}}(\mathbf{x}_{k+1}-\mathbf{x}_k)$
    \ENDFOR
    \ENSURE  Reconstructed signal $\hat{\mathbf{x}}$.
  \end{algorithmic}  
\end{algorithm}

The mathematical process of Block Convolutional FISTA (BC-FISTA) is summarized in Algorithm \ref{alg_convfista}. In practical implementations, the calculation of the gradient $\nabla f(\cdot)$ and the Lipschitz constant $L$ is contingent upon the explicit formulation of $f(\cdot)$. For instance, in scenarios where $f(\mathbf{x})$ represents a linear operation $f(\mathbf{x})$, the gradient is determined as $\mathbf{A}\trans(\mathbf{Ax-y})$, and $L$ can be identified as the largest eigenvalue of $\mathbf{A}\trans\mathbf{A}$. However, in the context of extensive ULAs and ROIs, the dimensions of the matrix $\mathbf{A}$ may surpass the available Random Access Memory (RAM) of CPUs or GPUs. This scenario can lead to suboptimal values of $L$ and $\lambda$, potentially resulting in slow convergence rates for the BC-FISTA. Therefore, in such cases, adopting a Learned Block Convolutional ISTA (BC-LISTA) emerges as the superior strategy. 

\subsection{Learned Block Convolutional ISTA}
Within the framework of deep algorithm unrolling \cite{monga2021algorithm}, the BC-LISTA requires a transpose model, denoted as $g(\cdot)$, for the computation of the back-projected residual. The mathematical formulation of the single-iteration update is presented as follows:
\begin{equation}
    \mathbf{x}_{k+1} = S_{\lambda/L}\Big(\mathbf{x}_k+\frac{1}{L}(g(\mathbf{y}-f(\mathbf{x}_k))) \Big).
\end{equation}

In the network architecture, $g(\cdot)$ is implemented with ConvTranspose1D layers. Each ConvTranspose1D layer is designed to deduce a reconstructed signal from an individual data slice. These reconstructed signals are subsequently aggregated through a weighted sum mechanism, endowed with trainable weights, to produce the refined signal estimate. This estimate is then processed through a soft-thresholding function, facilitating further updates.

The BC-LISTA architecture provides significant versatility in defining trainable parameters, including the kernels for the forward and transpose models, soft-thresholding parameters, and weights in the weighted sum mechanism. It is recommended to keep the forward model fixed to ensure consistency, while parameters such as $\lambda$, $L$, and aggregation weights are optimized as trainable to enhance network adaptability and performance. Typically, the kernels of the transpose model are adjustable to facilitate robust signal reconstruction, except in cases with optimal initial conditions or simpler inference tasks where fixed kernels may be adequate.

\subsection{Evaluation}
In the evaluation, we employ a $64$-channel ULA, adhering to the parameters establishment scheme for Fig. \ref{fig_storage_compare}, specifically, $N_{\rm{c}}=64$ and $d_{\rm{c}}=0.5\, \rm{mm}$. Our training process for BC-LISTA involved synthetic data generation; for each epoch, we produced a new set of random reflectivity maps $\{\mathbf{x}_n\}$ as in \cite{wangicassp2024}, and converted these into FMC data $\{\mathbf{Y}_n\}$, where $n \in [1, \dots, 20]$. The Adam optimizer with learning rate $1\times 10^{-4}$ is used to minimize the mean squared error $\frac{1}{N}\Vert \text{BC-LISTA}(\mathbf{Y}_n) - \mathbf{x}_n \Vert$ across $50$ epochs. The learned parameters are the final superposition weights, as well as the kernels of the transpose model $g(\cdot)$, the soft-thresholding parameter $\lambda$, and step size $L$ for each layer. 
 
\begin{figure}[ht]
    \centering
    \begin{subfigure}[b]{0.3\columnwidth}
        \centering
        \includegraphics[width=\linewidth]{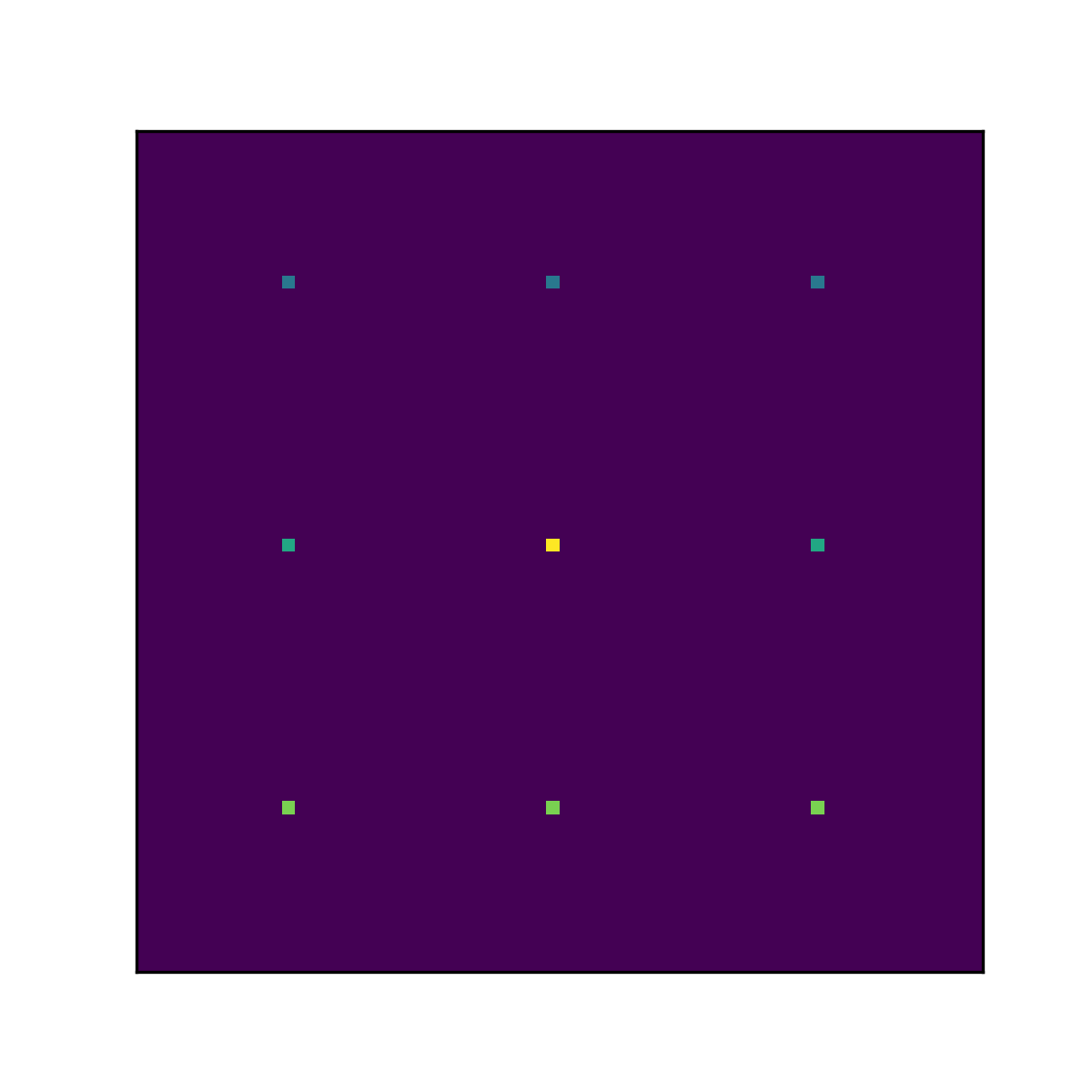}
        \caption{Ground Truth}
        \label{fig:sub1}
    \end{subfigure}
    \hfill
    \begin{subfigure}[b]{0.3\columnwidth}
        \centering
        \includegraphics[width=\linewidth]{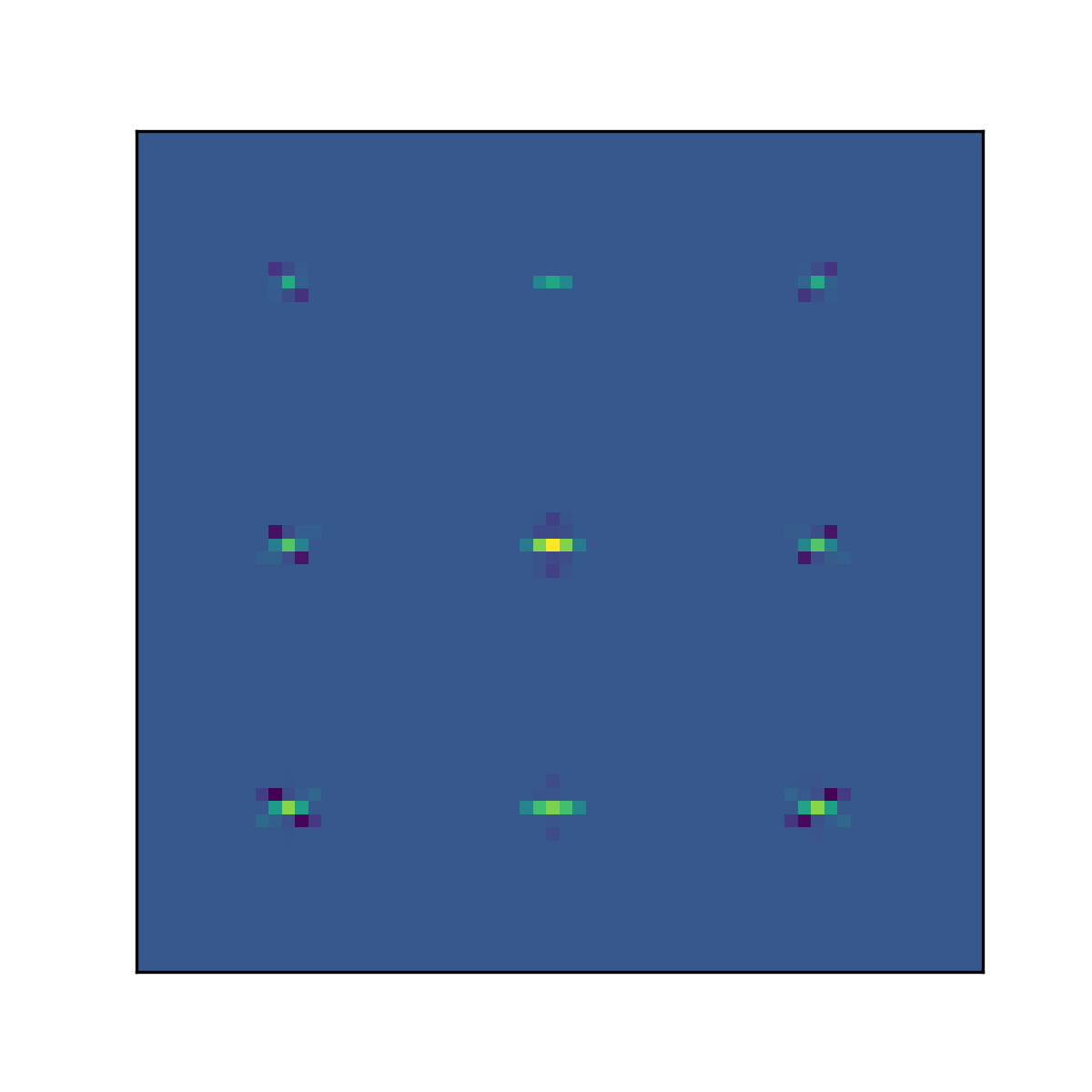}
        \caption{BC-LISTA}
        \label{fig:sub2}
    \end{subfigure}
    \hfill
    \begin{subfigure}[b]{0.3\columnwidth}
        \centering
        \includegraphics[width=\linewidth]{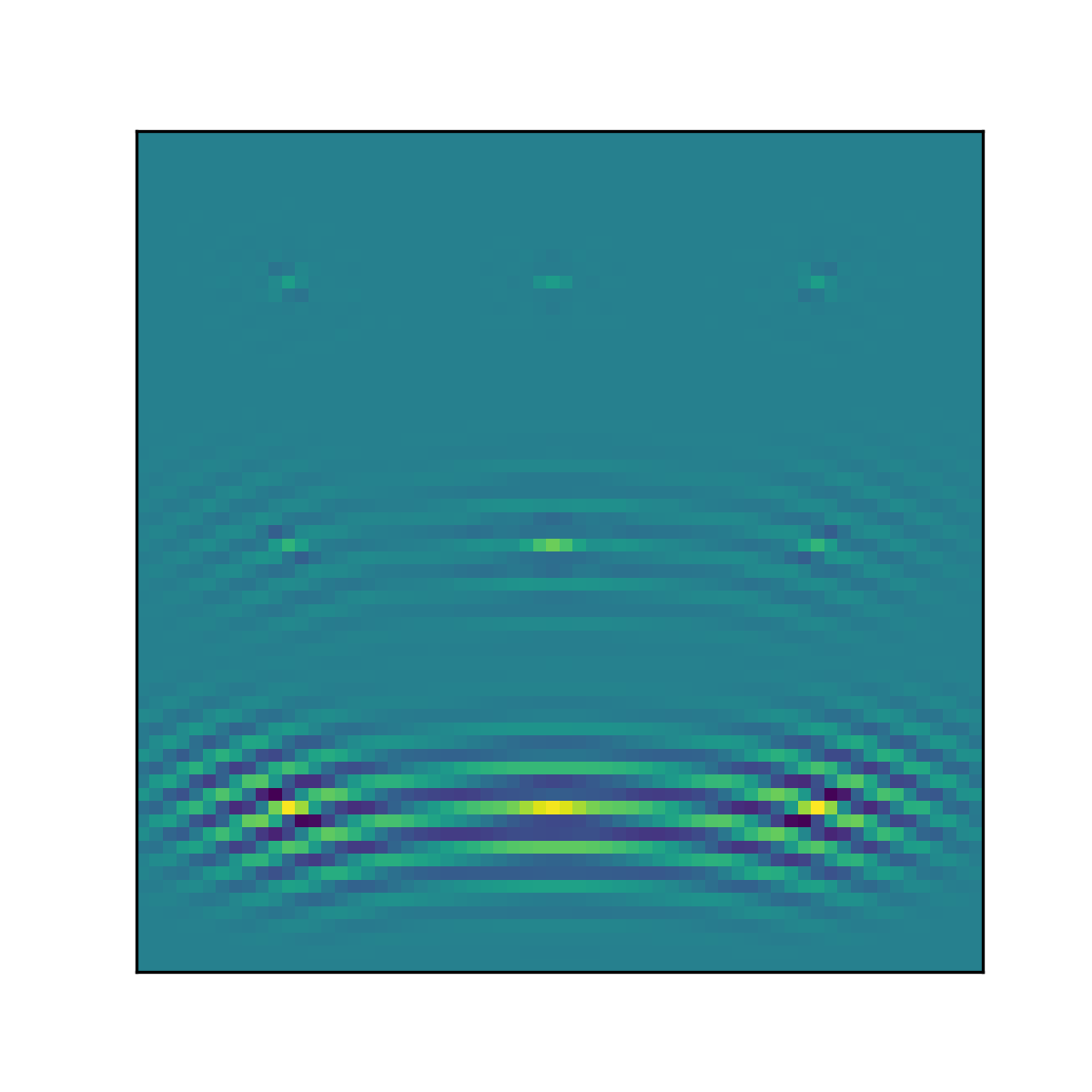}
        \caption{BC-FISTA}
        \label{fig:sub3}
    \end{subfigure}
    \caption{Reconstructed images comparison. The reconstruction achieved through BC-FISTA is mathematically identical to that of the naive FISTA algorithm but at a lower cost.}
    \vspace{-0.7cm}
    \label{fig_reco_compare}
\end{figure}

We deploy a $5$-layer BC-LISTA alongside a 5-iteration BC-FISTA, applied to identical FMC data sets. The comparative analysis of the resultant images is visually depicted in Fig. \ref{fig_reco_compare}, where it is evident that BC-LISTA significantly outperforms BC-FISTA in terms of image quality.

Subsequently, both algorithms were executed 100 times on the same dataset, with processing times recorded in seconds on both CPU and GPU platforms. The CPU tests were conducted on a laptop with an Intel Core i7-8550U processor, while the GPU tests were performed on a cluster equipped with AMD EPYC 7742 processors and NVIDIA A100 GPUs. This analysis yielded maximum, minimum, and average execution times, detailed in TABLE \ref{tab:computation-efficiency}. The findings highlight the computational efficiency of the CSC algorithms on both platforms, with BC-LISTA demonstrating superior performance on both platforms. Additionally, the rapid processing times achieved on the GPU indicate an efficient training process for BC-LISTA.

\begin{table}[ht]
\vspace{-0.15cm}
\centering
\caption{Computation Efficiency Comparison}
\label{tab:computation-efficiency}
\resizebox{\columnwidth}{!}{
\begin{tabular}{
  l 
  S[table-format=1.3] 
  S[table-format=1.3] 
  S[table-format=1.3] 
  S[table-format=1.3] 
  S[table-format=1.3] 
  S[table-format=1.3]
}
\toprule
& \multicolumn{3}{c}{Intel Core i7-8550U} & \multicolumn{3}{c}{NVIDIA A100} \\
\cmidrule(lr){2-4} \cmidrule(lr){5-7}
& {Max} & {Ave} & {Min} & {Max} & {Ave} & {Min} \\
\midrule
BC-FISTA &22.317  &19.136  &16.628  & 1.657 & 1.183 & 1.175 \\
BC-LISTA &8.769  &6.518  &5.316  & 0.885 & 0.884 & 0.883 \\
\bottomrule
\multicolumn{7}{r}{\small \textsuperscript{*}Unit: seconds (s)} \\
\end{tabular}
}
\vspace{-0.15cm}
\end{table}

Performance differences are attributed to BC-FISTA's reliance on automatic differentiation for inference, which is computationally more intensive than BC-LISTA's convolutions and transpose operations. BC-LISTA benefits from using original convolution kernels as initial guesses, requiring fewer training epochs. Consistently, BC-LISTA outperforms BC-FISTA with an equal number of iterations, due to the advantages of employing learned linear transformations. 

\section{Conclusion}\label{sec_conclusion}
In this investigation, we suggest an innovative convolutional forward modeling technique that capitalizes on the Block-Toeplitz structure inherent in multichannel, time-delay-based matrices. This methodology is particularly advantageous for applications exhibiting shift-invariance along at least one dimension, where the corresponding axes are integer multiples of one another. We have demonstrated that our convolutional model markedly improves memory efficiency, an essential feature for managing large ULAs and expansive ROIs. Moreover, we managed to implement and assess two CSC algorithms, BC-FISTA and BC-LISTA, which not only perform effectively on high-end GPUs but also on older-generation CPUs. Notably, BC-LISTA delivers superior image quality and computational efficiency over (BC-)FISTA, underscoring the potential to accelerate data processing tasks significantly. The adaptability of our proposed model extends to other unrolled inversion algorithms, offering remarkable flexibility in the customization of trainable parameters. The detection of sparsity in the kernels provides a foundation for future research aimed at devising a more streamlined learned convolutional forward model, potentially simplifying the architecture further and reducing the requisite parameters. 

\bibliographystyle{unsrt}
\bibliography{literature}

\end{document}